# Fragmentation in Large Object Repositories
## Experience Paper

Russell Sears
University of California, Berkeley
sears@cs.berkeley.edu

Catharine van Ingen
Microsoft Research
vanIngen@microsoft.com

## ABSTRACT

Fragmentation leads to unpredictable and degraded application performance. While these problems have been studied in detail for desktop filesystem workloads, this study examines newer systems such as scalable object stores and multimedia repositories. Such systems use a get/put interface to store objects. In principle, databases and filesystems can support such applications efficiently, allowing system designers to focus on complexity, deployment cost and manageability.

Although theoretical work proves that certain storage policies behave optimally for some workloads, these policies often behave poorly in practice. Most storage benchmarks focus on short-term behavior or do not measure fragmentation. We compare SQL Server to NTFS and find that fragmentation dominates performance when object sizes exceed 256KB-1MB. NTFS handles fragmentation better than SQL Server. Although the performance curves will vary with other systems and workloads, we expect the same interactions between fragmentation and free space to apply.

It is well-known that fragmentation is related to the percentage free space. We found that the ratio of free space to object size also impacts performance. Surprisingly, in both systems, storing objects of a single size causes fragmentation, and changing the size of write requests affects fragmentation. These problems could be addressed with simple changes to the filesystem and database interfaces. It is our hope that an improved understanding of fragmentation will lead to predictable storage systems that require less maintenance after deployment.

### Categories and Subject Descriptors
D.4.3 [**File Systems Management**], D.4.8 [**Performance**], H.3.2 [**Information Storage**], K.6.2 [**Installation Management**]

## 1. INTRODUCTION

Application data objects continue to increase in size. Furthermore, the increasing popularity of web services and other network applications means that systems that once managed static archives of "finished" objects now manage frequently modified versions of application data. Rather than updating these objects in place, typical archives either store multiple versions of the objects (the V of WebDAV stands for "versioning" [25]), or simply do wholesale replacement (as in SharePoint Team Services [19]). Similarly, applications such as personal video recorders and media subscription servers continuously allocate and delete large, transient objects.

Applications store large objects as some combination of files in the filesystem and as BLOBs (binary large objects) in a database. Only folklore is available regarding the tradeoffs. Most designers say that a database is probably best for small objects and that that files are best for large objects. However, the breakpoint is fuzzy and the underlying causes are unclear.

This article compares the performance of two different implementations of an abstracted write-intensive web application. The first implementation uses SQL Server to store the objects as BLOBs. The second uses NTFS to store the objects as files. We measure how performance changes over time as the storage becomes fragmented.

One surprising (to us) conclusion of our work is that storage fragmentation is the main determinant of the break-even point between the systems. In essence, the filesystem seems to have better fragmentation handling than the database and this drives the break-even point down from about 1MB to about 256KB.

By characterizing the fragmentation behavior of large object repositories, we raise a number of important questions that are left unaddressed: little is understood about the allocation patterns and object lifetimes of these applications, and our experiments reveal unexpected interactions between high-level interfaces and storage layout policies. The filesystem and database communities would benefit from a mutual understanding of existing techniques and fragmentation.

## 2. BACKGROUND

Filesystems have long used sophisticated allocation strategies to avoid fragmentation while laying out objects on disk. For example, OS/360's filesystem was extent based and clustered extents to improve access time. The VMS filesystem included similar optimizations and provided a file attribute that allowed users to request a (best effort) contiguous layout [9, 13]. Berkeley FFS [14] was an early UNIX filesystem that took sequential access, seek performance and other hardware characteristics into account. Subsequent filesystems were built with similar goals in mind.

The filesystem used in these experiments, NTFS, uses a 'banded' allocation strategy for metadata, but not for file contents [Microsoft NTFS Development Team, Personal Communication]. NTFS allocates space for file stream data from a run-based lookup cache. Runs of contiguous free clusters are ordered in decreasing size and volume offset. NTFS attempts to satisfy a new space allocation from the outer band. If that fails, large extents within the free space cache are used. If that fails, the file is fragmented. Additionally, the NTFS transactional log entry must be committed before freed space can be reallocated after file deletion. The net behavior is that file stream data tends to be allocated contiguously within a file.

Databases historically focused on small (100-byte) records and on clustering tuples within the same table. Clustered indexes let users control the order in which tuples are stored, allowing common queries to be serviced with a sequential scan over the data.

Filesystems and databases take different approaches to modifying an existing object. Filesystems are optimized for appending or truncating a file. When data are inserted or deleted in the middle of a file, such as when an element is added to an XML document, all contents after the modification must be completely rewritten. Some databases completely rewrite modified BLOBs; this rewrite is transparent to the application. Others, such as SQL Server, adopt the Exodus design that supports efficient insertion or deletion within an object by using B-Tree based storage of large objects [4]. In such systems, insertions and deletions within an object can lead to fragmentation [3].

Hybrid approaches exist as well. POSTGRES 4 can store BLOBs in the filesystem or in database rows, and provides a filesystem-style interface to applications. It also supports various compression schemes and server side processing of BLOBs [22]. IBM DB2's DataLinks technology stores BLOBs in the filesystem, and uses the database as an associative index of the file metadata [2]. Their files are updated atomically using a mechanism similar to safe writes (Section 4).

## 3. PRIOR WORK

We know of few systematic studies of fragmentation and little hard data on the actual performance of fragmented systems. A number of benchmarks are impacted by fragmentation, but do not measure fragmentation per se. There are many discussions of space allocation algorithms. However, existing studies do not cover recent changes in applications and the increasing popularity of databases.

### 3.1 Folklore

There is a wealth of anecdotal experience with applications that use large objects. The prevailing wisdom is that databases are better for small objects while filesystems are better for large objects. The boundary between small and large is a bit fuzzy. The usual reasoning is:

- Database queries are faster than file opens. The overhead of opening a file handle dominates performance for small objects.
- While file opens are CPU expensive, they are easily amortized over the cost of streaming large objects.
- Reading or writing large files is faster than accessing large database BLOBs. Filesystems are optimized for streaming large objects.
- Database client interfaces are not designed for large objects. Instead, they have been optimized for short low-latency requests returning small amounts of data.

None of the above points address the question of application complexity. Applications that store large objects in the filesystem encounter the question of how to keep the database object metadata and the filesystem object data synchronized. A common problem is the garbage collection of files that have been "deleted" in the database but not the filesystem. Also missing are operational issues such as replication, backup, disaster recovery, and fragmentation.

### 3.2 Theoretical results

The bulk of the theoretical work focuses on online schemes that always choose a contiguous layout for file data, but that may waste space. Early work bins small files into a set of disk tracks so that no file spans more than one track. This guarantees that each file can be read using a single disk seek. A variant of first fit always allocates space that is no more than 1.223 times larger than the optimal binning [6]. In the worst case, an optimal layout policy that allowed larger files, but still guaranteed that files are contiguous would use between $0.5\,M\log_2 n$ and $0.84\,M\log_2 n$ bytes of storage, where $M$ is the number of bytes of data, and $n$ is the maximum file length. First fit is nearly optimal, using at most $M\log_2 n$ bytes [16]. Unfortunately, for objects in the megabyte to gigabyte range, these bounds imply volumes should be roughly 20 to 30 times larger than the data they store.

Later work shows that, given some simplifying assumptions, some layout policies have acceptable average case performance with certain workloads [5]. Allocation algorithms that fragment objects are not considered; fragmentation arises only from space between objects. Also, inter-file locality and seek times are not considered.

The question of disk space layout is both similar to and different from memory allocation (malloc). In both cases, space can become fragmented and similar simplifying assumptions are often made. Disk space layout differs from memory layout in that the cost of a poor layout is larger—a disk seek rather than a cache miss—and that the cost of repairing a very fragmented disk is larger—a backup restore or total recreation of a database on new storage rather than a process restart or system reboot.

Borrowing from the malloc literature [26], we distinguish between allocation mechanisms, such as the buddy system, boundary tags and B-trees, and the policies that they approximate, such as best fit, and preserving locality of consecutive requests.

In practice, deallocation patterns can be highly structured. For example, pictures shared for an event are often uploaded and later deleted as a group. Ideally, such deallocation would produce large, contiguous regions of free space. Empirical studies have shown that objects malloced at the same time are usually freed at the same time. Therefore, using a large, contiguous region for a collection of related allocations tends to preserve the contiguous region for eventual reuse.

Until recently, it was standard practice for theoretical studies to assume objects are deallocated randomly. This allows algorithms that ignore structure in file deletion requests to perform well on synthetic workloads, and explains why theoretically optimal allocation strategies can perform poorly in practice.

Applications that concurrently process unrelated requests complicate the situation because temporal clustering is no longer adequate [1]. Web services encounter multiple groups of users, each of which exhibit different behaviors. For example, the allocation patterns of ad hoc sharing and long-term archival differ, and members of flash crowds behave in different ways than loyal web site visitors.

Applications that partition data based upon object sizes, expected workloads and deallocation patterns, or that use underlying storage in degenerate ways (Section 5.4) can also complicate matters. Such behavior may destroy allocation patterns or hide them from the storage layer.

As drive capacity increases, it may be worthwhile to trade capacity for predictability and implement persistent storage systems that do not support file fragmentation. For these reasons, the relevance of early theoretical results may increase over time.

### 3.3 Standard benchmarks

Most filesystem benchmarking tools consider only the performance of clean filesystems and do not evaluate long-term performance as the storage system ages and fragments. Using simple initial conditions eliminates potential variation in results and reduces the need for settling time to allow the system to reach equilibrium.

Several long-term filesystem fragmentation and performance studies have been performed based upon two general approaches [18]. *Trace based* load generation uses data gathered from production systems over a long period. *Vector based* load generation models application behavior as a list of primitives and randomly applies each primitive with the frequency of a real application.

NetBench [15] is the most common Windows file server benchmark. It measures the performance of a file server accessed by multiple clients using office applications.

SPC-2 benchmarks storage system applications that read and write large files in place, execute large read-only database queries, or provide read-only on-demand access to video files [21].

The Transaction Processing Performance Council [24] defined several benchmark suites to characterize online transaction processing workloads and decision support workloads. These benchmarks do not address large objects or multimedia databases.

None of these benchmarks explicitly consider file fragmentation.

### 3.4 Data layout approaches

The creators of FFS observed that for typical workloads of the time, fragmentation avoiding allocation algorithms kept fragmentation under control as long as volumes were kept under 90% full [20]. UNIX variants still reserve a certain amount of free space on the drive, both for disaster recovery and in order to prevent excess fragmentation.

The Dartmouth Time-Sharing System (DTSS) file system was developed in the 1960's and used by the Honeywell DPS/8. It lays out files using the buddy system, which imposes hard limits on the number of fragments that can be used to store files of various lengths. Although it had good fragmentation behavior, the fragmentation limits it imposed were problematic for applications that created large files [12].

NTFS disk occupancy on deployed Windows systems varies widely. System administrators target disk occupancy as low as 60% or over 90% [Microsoft NTFS Development Team, Personal Communication]. The Windows defragmentation utility supports on-line partial defragmentation including system files.

LFS [17], a log-based filesystem, optimizes for write performance by organizing data on disk according to the chronological order of the write requests. This allows it to service write requests sequentially but causes severe fragmentation when files are updated randomly. A cleaner that simultaneously defragments the disk and reclaims deleted file space can partially address this problem.

XFS [23] is a filesystem designed for large-object storage that uses delayed allocation and best fit to avoid fragmentation. Later versions of FFS adopt a similar policy, *realloc*, which groups logically sequential blocks into physically contiguous clusters as they are flushed to disk. The maximum cluster length was typically set to the storage system's maximum transfer size. The realloc allocation policy roughly halved the disk fragmentation caused by a trace-based workload generator [20]. Delayed allocation was later adopted by other filesystems such as EXT2 and ReiserFS.

Network Appliance's WAFL ("Write Anywhere File Layout") [11] is able to switch between conventional and write-optimized file layouts depending on workload conditions. WAFL also leverages NVRAM caching for efficiency and provides access to snapshots of older versions of the filesystem contents. Rather than a direct copy-on-write of the data, WAFL metadata remaps the file blocks. A defragmentation utility is supported but is said not to be needed until disk occupancy exceeds 90+%.

While most approaches to fragmentation focus on the number of seeks required to read a file, newer drives are divided into zones that transfer data at different bandwidths. An optimal policy for placing popular files in faster zones has been developed, along with online reorganization to defragment and migrate popular files. Simulations show a 20-40% performance improvement on FTP workloads [7]. NTFS's banded allocation strategy is designed to take advantage of disk zones. Similarly, its defragmenter moves application and boot files to faster bands.

GFS [8], a filesystem designed to deal with multi-gigabyte files on 100+ terabyte volumes, addresses the data layout problem by using 64MB *chunks*. GFS provides a safe record append operation that allows multiple simultaneous client appends to the same file, reducing the number of files (and opportunities for fragmentation) exposed to the underlying filesystem. GFS records may not span chunks, which can result in internal fragmentation. If a record will not fit into the current chunk, that chunk is zero padded, and a new chunk is allocated. Records are constrained to be less than ¼ the chunk size to prevent excessive internal fragmentation. However, GFS does not explicitly attempt to address fragmentation introduced by the underlying filesystem, or to reduce internal fragmentation after records are allocated.

## 4. COMPARING FILES AND BLOBS

Our experiments focused on applications that make use of simple get/put storage primitives. Such applications are quite common and include most desktop applications such as word processors, collaboration tools such as SharePoint, applications built on top of WebDAV, web services such as photo sharing, some web mail implementations, and map and GIS services.

To present a fair comparison of NTFS and SQL Server, we were careful to ensure that they provide similar semantics. Under

NTFS, we use *safe writes* to atomically update objects. To perform a safe write an application writes the object to a temporary file, forces that file to be written to disk, and then atomically replaces the permanent file with the temporary file. Typically, this is done using ReplaceFile() under Windows, or rename() under UNIX.

Under SQL Server, we used *bulk logged* mode. In this mode, newly allocated BLOBs are written to the page file and forced to disk at commit. This avoids the log write while providing normal ACID transactional semantics. In this way, neither NTFS nor SQL Server support BLOB recovery after media failure as there is no second copy of the BLOB.

For simplicity, we did not consider the overhead of detecting and correcting silent data corruption and media failures. These overheads are comparable for file and database systems, and typically involve maintenance of multiple copies with periodic "scrubbing" to detect and correct data corruption.

Since our purpose was to fairly evaluate the out-of-the-box performance of the two storage systems, we did no performance tuning except in cases where the default settings introduced gross discrepancies in the functionality that the two systems provided.

### 4.1 File based storage

Following the practice of our applications, we stored object names and other metadata in SQL server tables. Each application object was stored in its own file. The files were placed in a single directory on an otherwise empty NTFS volume. SQL was given a dedicated log and data drive, and the NTFS volume was accessed via UNC path.

This partitioning is fairly flexible and allows a number of replication and load-balancing schemes. The database isolates the client from the physical location of data—changing the pointer in the database changes the path returned to the client.

We chose to measure performance with the database co-located with the associated files. This configuration kept our experiments simple and independent of the network layout. However, we structured all code to use the same interfaces and services as a networked configuration.

### 4.2 Database storage

The database storage tests were designed to be as similar to the filesystem tests as possible. We stored the BLOBs and the metadata in the same filegroup, but we used out-of-row storage for the BLOB data so that the BLOBs did not decluster the metadata. Out-of-row storage places BLOB data on pages that do not store other table fields, allowing the table data to be kept in cache.

Analogous table schemas and indices were used and only minimal changes were made to the software that performed the tests.

### 4.3 Workload generation

The applications of interest are extremely simple from the storage point of view. A series of object allocation, deletion, and safe-write updates are processed with interleaved read requests.

For simplicity, we assumed that all objects are equally likely to be written and/or read. We also assumed that there is no correlation among objects. Recall from Section 3.2 that this simple allocation model is inappropriate for any structured real-world workload.

We measured constant size objects rather than objects with more complicated size distributions. We expected size distribution to be an important factor in our experiments. As discussed below, we found that it had no obvious effect on fragmentation behavior.

We considered more sophisticated synthetic workloads, but concluded that the added complexity would not lead to more insight. First, given current understanding of large object applications, high-level caches, load balancing and heat-based object partitioning, any distribution we chose would be based on speculation. Second, complex workloads often obfuscate simple yet important effects by making it difficult to interpret the results.

### 4.4 Storage age

Past fragmentation studies measure age in elapsed time such as days or months [20], or in the total amount of work performed [3]. However, elapsed time is not very meaningful when applied to synthetic traces, and measurements of total work performed depend on volume size. We considered reporting age in "hours under load", but this would allow slow storage systems to perform less work during the test.

We measure time using *storage age*; the ratio of bytes in objects that once existed on a volume to the number of bytes in use on the volume. This definition assumes that the amount of free space on a volume is relatively constant over time but allows comparison of workloads with different distributions of object lifetimes. Storage age is similar to the VIFS performance study's *generations*; a new generation begins when each file from the previous generation has been deleted and replaced [10].

In a safe-write system, storage age is the ratio of object insert-update-delete bytes divided by the number of live object bytes. For our workload, this is "safe writes per object".

Given an application trace, storage age can be computed from the data allocation rate. This allows synthetic workloads to be compared to trace-based workloads. Storage age is independent of volume size and update strategy. Therefore it can be compared across hardware configurations and applications.

## 5. RESULTS

We used throughput as the primary indicator of performance, and started by benchmarking a clean system. We then looked at the longer-term changes caused by fragmentation with a focus on 256K to 1M object sizes where the filesystem and database have comparable performance. Finally, we discuss the effects of volume size and object size distribution.

### 5.1 Test system configuration

All the tests were performed on the system described in Table 1. All test code was written using C# in Visual Studio 2005 Beta 2 and was compiled to x86 code with debugging disabled.

**Table 1. Configuration of the test system**

| |
|---|
| Tyan S2882 K8S Motherboard, 1.8 Ghz Opteron 244, 2 GB RAM (ECC) |
| SuperMicro "Marvell" MV8 SATA controller |
| 4 Seagate 400GB ST3400832AS 7200 rpm SATA drives |
| Windows Server 2003 R2 Beta (32 bit mode) |
| SQL Server 2005 Beta 2 (32 bit mode) |

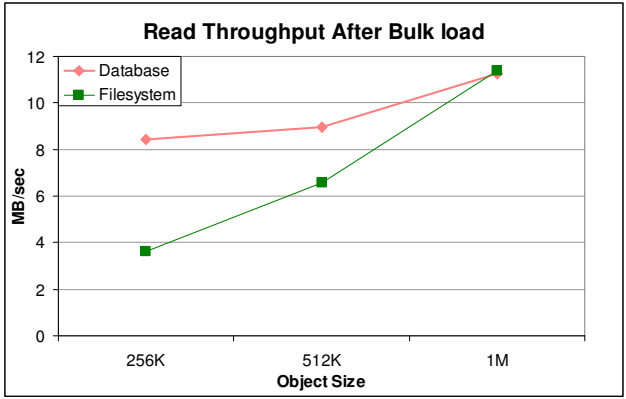

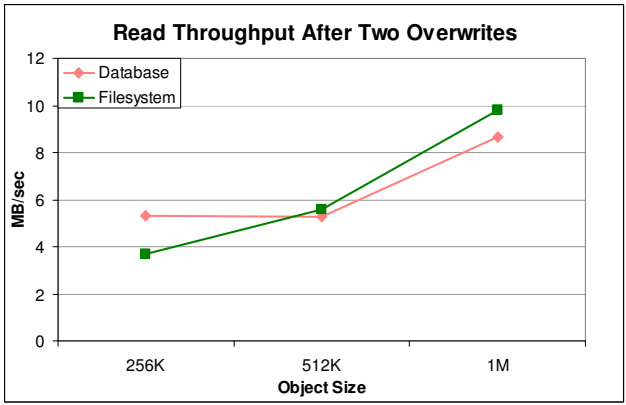

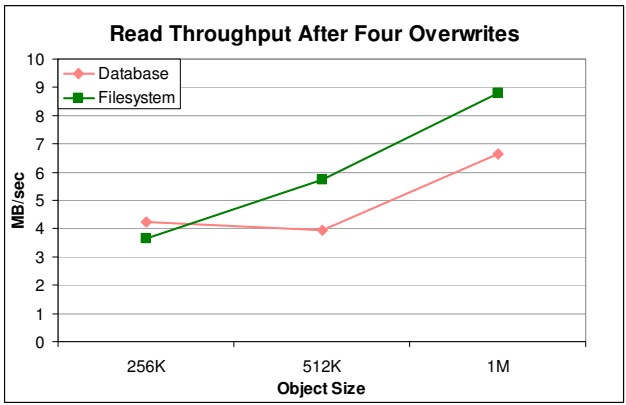

**Figure 1. Read throughput. Immediately after bulk load, SQL Server is faster on small objects while NTFS is faster for large objects. As objects are overwritten, fragmentation degrades SQL Server's performance. NTFS eventually outperforms SQL Server on objects greater than 256KB.**

## 5.2 Initial throughput

We begin by establishing when a database is clearly the right answer and when the filesystem is clearly the right answer. On a clean data store, Figure 1 demonstrates the truth of the folklore: objects up to about 1MB are best stored as database BLOBs. With 10MB objects, NTFS outperforms SQL Server. Interestingly, the write throughput of SQL Server exceeded that of NTFS during bulk load. With 512KB objects, database write throughput was 17.7MB/s, while the filesystem only achieved 10.1MB/s (Figure 4).

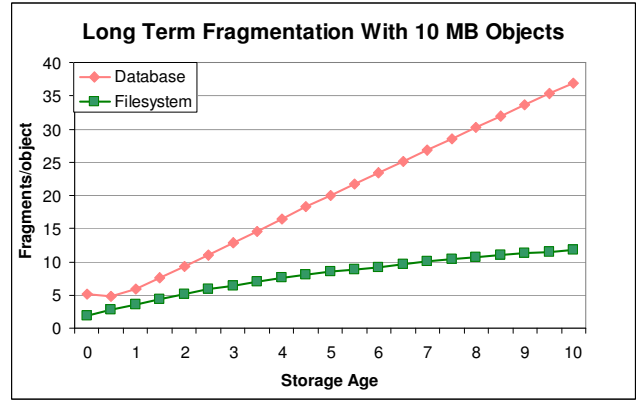

**Figure 2. Large object fragmentation. "Storage Age" is the ratio of deleted objects to live objects. Contiguous objects have 1 fragment.**

## 5.3 Performance over time

Next, we evaluate the performance over time. If fragmentation is important, we expect to see noticeable performance degradation.

Our first discovery was that SQL Server's fragmentation reports and defragmentation tools handle index data but not large object data. The recommended way to defragment a large BLOB table is to create a new table in a new file group, copy the old records to the new table and drop the old table [Microsoft SQL Server Development Team, Personal communication].

To measure fragmentation, we tagged each of our objects with a unique identifier and a sequence number at 1KB intervals, and then determined the physical locations of these markers on the hard disk. We validated this tool against the Windows NTFS defragmentation utility by comparing the reports.

The degradation in read performance for 256K, 512K, and 1MB BLOBs is shown in Figure 1. Each storage age (2 and 4) corresponds to the time necessary for the number of updates, inserts, or deletes to be N times the number of objects in our store since the bulk load (storage age 0).

For large objects, fragmentation under NTFS begins to level off over time, while SQL Server's fragmentation increases almost linearly over time and does not seem to be approaching any asymptote (Figure 2). When we ran on an artificially and pathologically fragmented NTFS volume, we found that fragmentation slowly decreases over time. This suggests that NTFS is indeed approaching an asymptote.

These results indicate that as storage age increases, the break-even point where NTFS and SQL Server have comparable performance declines from 1MB to 256KB. Within this range, fragmentation eventually halves SQL Server's throughput. Objects up to about 256KB are best kept in the database; larger objects should be in the filesystem.

To verify this, we attempted to run both systems until their performance converged to comparable steady states. Figure 3 indicates that fragmentation in both systems converges to four fragments per file, or one fragment per 64KB. Our tests use 64KB write requests, suggesting that the impact of the size of file creation and append operations upon fragmentation warrants further study. From this data, we conclude that SQL Server's read performance is indeed superior to NTFS on objects under 256KB.

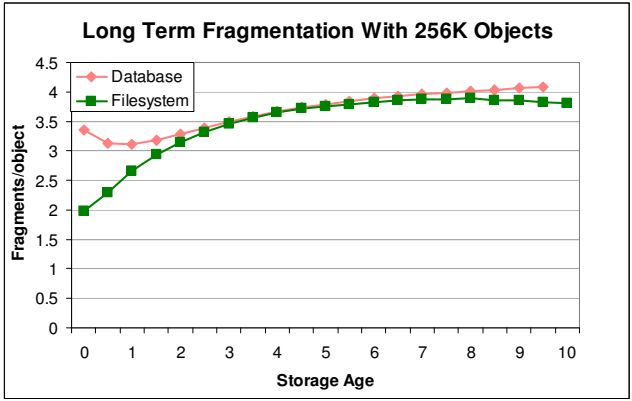

**Figure 3. For small objects, the systems have similar fragmentation behavior.**

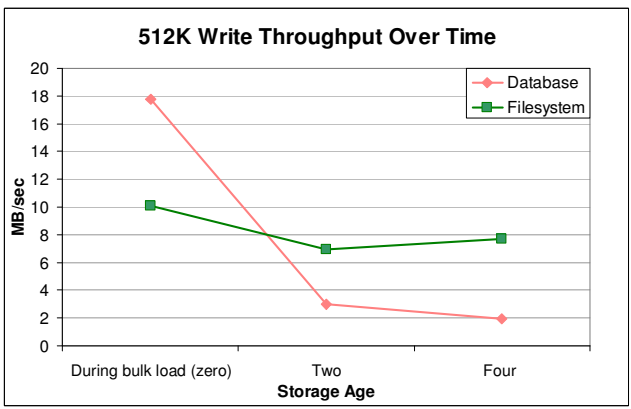

**Figure 4. Although SQL Server quickly fills a volume with data, its performance suffers when existing objects are replaced.**

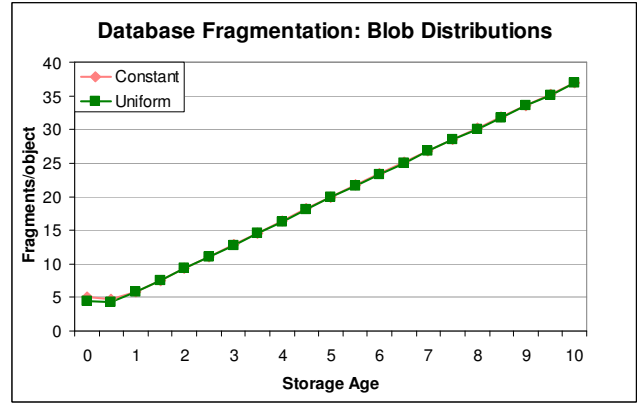

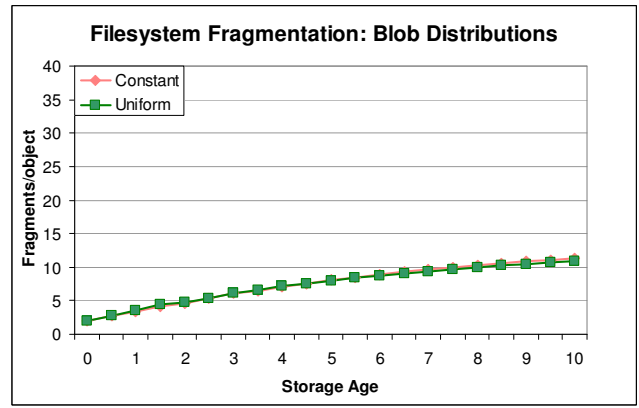

**Figure 5. Fragmentation for large (10MB) BLOBs increases slowly for NTFS but rapidly for SQL. Surprisingly, objects of a constant size show no better fragmentation performance than objects of sizes chosen uniformly at random with the same average size.**

Figure 4 shows the degradation in write performance as storage age increases. In both systems, the write throughput during bulk load is much better than read throughput immediately afterward. Both can simply append each new object to the end of allocated storage, avoiding seeks. On the other hand, the read requests are randomized and incur at least one seek. As the fragmentation behavior of the two systems suggests, the write performance of SQL Server degrades quickly after bulk load.

Note that the read and write performance numbers are not directly comparable. The read performance is measured after fragmentation, while write performance is measured during fragmentation. For example, the "storage age two" write performance is the average write throughput between the "bulk load" and "storage age two" read measurements.

## 5.4 Fragmentation effects of object size, volume capacity, and write request size

Distributions of object size vary greatly. Similarly, applications are deployed on storage volumes of widely varying size. This section describes the effects of varying object size distributions and volume sizes. All experiments presented here make use of objects with a mean size of 10MB.

Our intuition suggested that constant size objects should not fragment. Best fit, first fit and worst fit all behave optimally on this workload—deleting a contiguous object always leaves a region of contiguous free space exactly the right size to store any other object. To confirm this, we compared fragmentation behavior with a constant object size and with object sizes drawn from a uniform distribution. As shown in Figure 5, our intuition was wrong. As long as the average object size is held constant there is little difference—fragmentation is occurring with both constant- and uniform-sized objects.

We believe the fragmentation occurs because NTFS allocates space as the file is being appended to, which is before it knows the final size. Also, modifying the size of the write requests that append to NTFS files and database BLOB's changes long-term fragmentation behavior, supporting this theory.

There is no way to pass the (known) object size to the file system at file creation and initial space allocation. NTFS will aggressively attempt to allocate contiguous space when sequential appends are detected, but there is no guarantee of contiguous allocation. Systems that use deferred allocation partially address this problem by implicitly increasing the size of file append requests. These systems trade system memory to buffer write requests for improved information about the object's final size. Although changing the allocation interface would improve performance for get/put applications, it is unclear what fraction of applications these represent.

The time it takes to run the experiments is proportional to the volume's capacity. When the entire disk capacity (400GB) is used, some experiments take a week to complete. Using a smaller (although perhaps unrealistic) volume size allows more experiments; but how trustworthy are the results?

As shown in Figure 6, the exact size of large (40-400GB) volumes has a relatively minor effect on performance. However, on smaller volumes, we found that as the *ratio of free space to object size* decreases, performance degrades. We did not characterize the exact point where this becomes a significant issue. Although larger volumes tend to outperform smaller volumes, our results suggest that the effect is negligible when there is 10% free space on a 40GB volume storing 10MB objects, or a pool of 400 free objects. On a 4GB volume with a pool of 40 free objects, performance degraded rapidly. We found that NTFS is able to take advantage of extremely large pools of free objects on volumes with low occupancy. When we keep the volumes 50% full, 400GB NTFS volumes converge to 4-5 fragments/object, while 40GB volumes converge to 11-12 fragments/object.

Our results suggest that extremely simple size distributions can be representative of many different get/put workloads. Our simple synthetic workload led to easily interpretable system behaviors and provided insight that would be difficult to obtain from trace-based studies. However, synthetic workloads are unrealistic models of real systems. Trace-based workload generation and a better understanding of real-world large object workloads would complement this study by allowing for realistic comparisons of large object storage techniques.

## 6. CONCLUSIONS

When designing a new system, it is important to consider behavior over time instead of looking only at the performance of a clean system. When fragmentation is a significant concern, the system must be defragmented regularly. However, defragmentation may require additional application logic and imposes read/write performance impacts that can outweigh its benefits.

Using storage age to measure time aids in the comparison of different designs. Here, we use safe-writes per object. In other applications, appends per object or some combination of create/append/deletes may be more appropriate.

For objects larger than 1MB, NTFS has a clear advantage over SQL Server. If the objects are smaller than 256 KB, the database has a clear advantage. Between 256KB and 1MB, storage age determines which system performs better.

The ability to specify the size of the object before initial space allocation could reduce fragmentation. We did not consider object update interfaces that allow arbitrary insertion and deletion of BLOB ranges; these would lead to different fragmentation behavior and application strategies. Also not considered were interleaved append requests to multiple objects, which are likely to increase fragmentation.

Information regarding the performance of fragmented storage is scarce, and characterization of other systems will likely provide additional insight. While both SQL Server and NTFS can be improved, the impact of such improvements is difficult to gauge without a better understanding of real-world deployments.

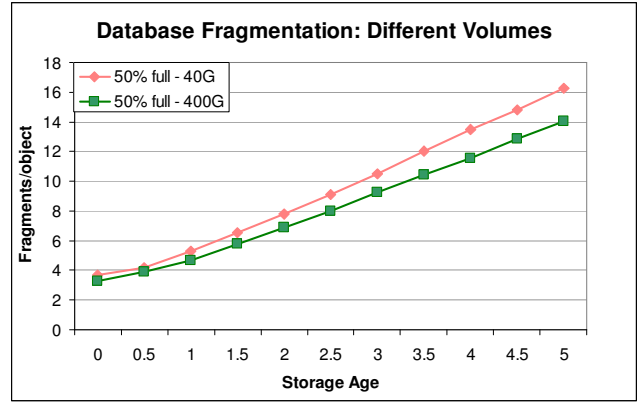

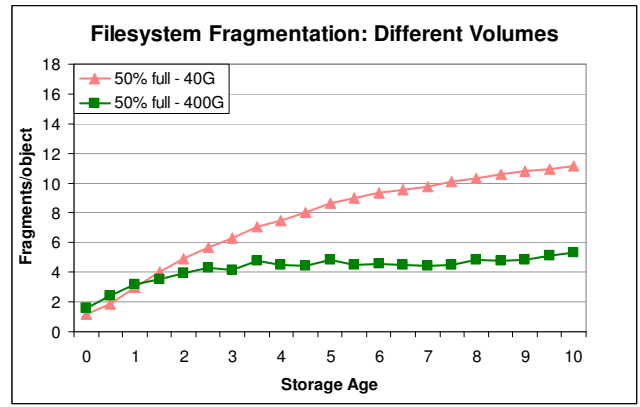

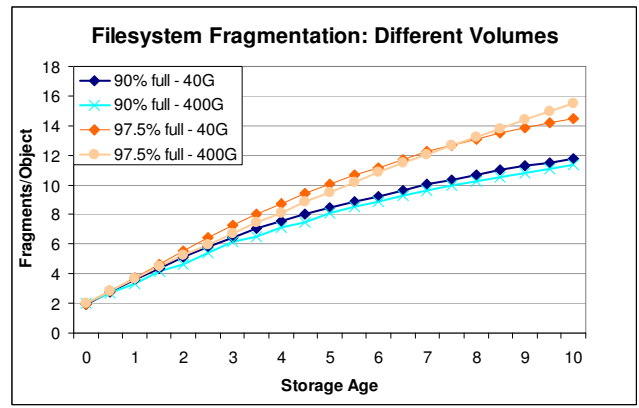

**Figure 6. Fragmentation for 40GB and 400GB volumes. Other than the 50% full filesystem run, volume size has little impact on fragmentation.**

## 7. ACKNOWLEDGMENTS

Many thanks to Jim Gray for suggestions that focused this study. Eric Brewer suggested the approach we used for the fragmentation analysis tool. Surendra Verma, Michael Zwilling and their development teams patiently answered questions throughout the study. We thank Dave DeWitt, Ajay Kalhan, Norbert Kusters, Wei Xiao, the anonymous reviewers, and all of the above for their constructive criticism.